\documentstyle[12pt]{article}

\protect
\protect
\newcommand{\lgm}{{\,\rm ln }}
\newcommand{\Break}{ \right. \nonumber \\ &{}& \left. }

\newcommand{\ice}[1]{\relax}
\newcommand{\prd}{\partial}
\newcommand{\ep}{\epsilon}
\newcommand{\beq}{\begin{equation}}
\newcommand{\eeq}{\end{equation}}
\newcommand{\bea}{\begin{eqnarray}}
\newcommand{\eea}{\end{eqnarray}}

\newcommand{\ba}{\begin{array}} 
\newcommand{\ea}{\end{array}} 
 
\newcommand{\as}{\alpha_s}

\newcommand{\G}{\Gamma}
\newcommand{\g}{\gamma}

\newcommand{\dmu}{\mu^2\frac{d}{d\mu^2}}
\newcommand{\msbar}{\overline{\mbox{MS}}}
\newcommand{\dsp}{\displaystyle}

\newcommand{\ovl}{\overline}

\begin{document}  

\begin{titlepage}

\begin{flushright}
\begin{tabular}{l}
  MPI/PhT/96-83\\
  hep-ph/9608480\\
  August   1996   
\end{tabular}
\end{flushright}
\mbox{}
\protect\vspace*{.3cm}

\begin{center}
  \begin{Large}
  \begin{bf}
Corrections of order $\as^3$ to $R_{had}$ 
in pQCD with light gluinos
 \\ \end{bf} \end{Large} \vspace{1cm}
K.G.~Chetyrkin $^{a,b}$,
\begin{itemize}
\item[$^a$]
 Institute for Nuclear Research,   
 Russian Academy of Sciences,   \\
 60th October Anniversary Prospect 7a 
 Moscow 117312, Russia 
\item[$^b$]{%
Max-Planck-Institut f\"ur Physik, Werner-Heisenberg-Institut, \\ 
F\"ohringer Ring 6, 80805 Munich, Germany}
\end{itemize}
\vspace{0.2cm}

  \vspace{0.5cm}
  {\bf Abstract}
\end{center}
\begin{quotation}
\noindent
By a completely independent analytic calculation made in general
covariant gauge we confirm the known result for the massless 
${\cal O}(\as^3)$ 
correction to  $R_{had}$
and check explicitly its gauge independence.  We extend the
calculations to include contributions due to a hypothetical existence
of a colour octet of (light) neutral fermions known also under the
name of gluinos.  In numerical form and for five active flavours 
we found  (in $\ovl{\mbox{MS}}$ scheme)     that 
$
R(s) = \frac{11}{3}[1 + \alpha_s/\pi + (1.409 - 0.346 n_{\tilde g}) (\alpha_s/\pi)^2                        
+(-12.805 - 3.006 n_{\tilde g} - 0.0466 n_{\tilde g}^2)(\alpha_s/\pi)^3
                     ]
$
\ice{                     
Out[26]= 1. + as + as^2  (1.40923 - 0.345886 ng) + 

>    as^3(-12.8046 - 3.00588 ng - 0.0466052 ng )
 
                     2
Out[22]= 1. + as + as  (1.409230409 - 0.3458861937 ng) + 
 
       3                                                  2
>    as  (-12.80462653 - 3.005884971 ng - 0.04660519788 ng )
    }
where  $n_{\tilde g}$ is   the number of  such colour octets. 
\end{quotation}
\vfill

\noindent
e-mail:
\\
chet@mppmu.mpg.de
\\
\end{titlepage}

\section{Introduction}
The simplest inclusive reaction involving quarks---their production
through a decay of a heavy virtual photon or a Z boson---is a process
of fundamental importance for QCD as the theory  of strong
interactions. Firstly, it provides us with a beautiful confirmation of
the existence of colour \cite{colour}. Secondly, the precision
measurements of $\G(Z \to \mbox{hadrons})$ have developed into an
important experimental tool for a reliable determination of $\as(M_Z)$
\cite{Blondel}. Thirdly, by comparing $\as(M_Z)$ with the value
of $\as$ as obtained from the measurements of $R(s)$ at lower energies
one could in principle directly test the evolution of the strong
coupling constant \cite{ChetKuhn95}.

Impressive wealth of theoretical results about $R(s)$ is now available
in the QCD framework (see e.g. a recent review \cite{review} for a
detailed discussion). 
In the massless  approximation, valid in the high energy limit, 
 corrections to
$R  \equiv \sigma(e^+ e^- \to \mbox{hadrons})/
           \sigma(e^+ e^- \to \mu^+ \mu^-)$
have been calculated up to order $\alpha_s^3$
\cite{GorKatLar91,SurSam91}.
For precision measurements the dominant mass corrections can  be
included through an expansion in $m_q^2/s$. Terms up to order
$\alpha_s^3 m^2_q/s$
\cite{CheKue90} 
and $\alpha_s^2 m^4_q/s^2$
\cite{CheKue94}
are available at present,
providing an acceptable approximation from the high energy
region down to intermediate energy values.
At order $\as^2$ even full account of the quark mass effects has been
recently made  on the basis of a semi-analytic approach
\cite{CheKueSte96Pade}.  Among massless calculations the one in the
order $\as^3$ is  probably  most involved. Let us discuss it
separately.

The results of the reevaluation of massless
next-next-to-leading $\as^3$ correction to $R(s)$ (first computed
erroneously in Ref.~\cite{GorKatLar88} because of   bugs  in a
computer program) were
published in Refs.~\cite{GorKatLar91,SurSam91} five years ago.  
Both calculations have produced identical
results and are much alike in many respects:  
\begin{itemize}
\item   the same  theoretical tools as well as computer programs 
       have been  used;
\item  the simplest gauge condition  -- the Feynman one -- has been employed;
\item An important subclass of all diagrams (which  includes, in fact, the
      most complicated ones) -- the QED type diagrams (that is those 
      not containing three 
      and four-gluon vertices) -- had been  computed  first  and 
      published in Ref.~\cite{qed-4loop} 
      by a collaboration   of authors of Refs.~\cite{GorKatLar91,SurSam91}. 
\end{itemize}
A golden rule well-known among multi-loop people says that a result of
a multi-loop calculation can be trusted and considered as {\bf the
result} only if it is confirmed by an independent calculation
preferably made by a different group and with the use of the general
covariant gauge.  Therefore, in view of the importance of 
the ${\cal O} (\alpha_s^3)$ correction 
for both theory and phenomenology it is necessary to
check those by a really independent calculation.

In the present paper we report about such an attempt.  Our work
confirms the results of Refs.~\cite{GorKatLar91,SurSam91} and
explicitly demonstrate their gauge independence. We have also computed
 extra diagrams which include virtual loops of 
Majorana fermions transforming as an octet with  respect to  the colour $SU_c(3)$
group.
Note that the case of Majorana fermions considered as {\em gluinos},
that is as superpartners for gluons \cite{Farrar78}, has some
phenomenological relevance at present. On one side, light gluino are
not completely excluded by the current experimental data
\cite{Ruckl94,Farrar96,Blumlein94}.  
On the other side, overall consistency of various  determinations of
$\as(M_Z)$ seems to improve if the light gluino with mass of a few GeV
does exist \cite{Jezabek93,Claveli9293,Ellis93}.

\section{Preliminaries}
From purely theoretical point of view $R(s)$ is an extremely suitable object
to deal with  within pQCD. This is because all  relevant information
is  contained in the current correlation function
\begin{equation}
\label{Pi}
\Pi_{\mu\nu}(q)  =
(4\pi)^2 i \int {\rm d} x e^{iqx}
\langle 0|T[ \;
\;j_{\mu}^{\rm em}(x)j_{\nu}^{\rm em}(0)\;]|0 \rangle
= 
\displaystyle
(-g_{\mu\nu}q^2  + q_{\mu}q_{\nu} )\Pi(-q^2)
{}\, ,
\end{equation}
with the hadronic EM current
$
j^{\rm em}_{\mu}=\sum_{{f}} Q_{{f}}
\overline{\psi}_{{f}} \gamma_{\mu} \psi_f
$, and $Q_f$ being the EM charge of the quark $f$.
The optical theorem
relates  the inclusive cross-section
and thus the function $R(s)$
to the discontinuity of $\Pi$
in the complex plane
\begin{equation} \label{d3}
R(s) =  \displaystyle
 \frac{3}{4\pi} \,{\rm Im}\, \Pi( - s -i\delta)
\label{discontinuity}
{}\, .
\end{equation}
\ice{
The polarization  operator $\Pi(Q^2)$ is related
through the standard dispersion relation to the ratio 
$R(s)$
\beq
\Pi (Q^2) = \frac{1}{12\pi^2}
\int_{0}^{\infty}ds
\frac{R(s}{s+Q^2}
 \;\;{\rm mod \;sub}
\label{dispersion.rel}
\eeq
     }  
Experimental ${\rm e}^+{\rm e}^-$ data are taken
in the physical regime of timelike
momentum transfer $q^2>0$. This region is influenced
by threshold and bound state effects which make
the use of perturbative QCD questionable.
However,
perturbative QCD is
strictly applicable for large
 spacelike momenta ($q^2=-Q^2<0$),
 since this region is far away
from non-perturbative
effects due to
hadron thresholds and
resonance effects \cite{Adl74}.
Therefore,
reliable theoretical predictions can
be made for $\Pi(Q^2)$ with $Q^2>0$.
To compare theoretical predictions and
experimental results for time-like momenta,
one has to perform suitable
averaging procedures \cite{PogQuiWei76,Chet78}.
For large positive ${s}$ one may appeal to the
experimentally observed smoothness of $R$
as a function of ${s}$ and to the absence
of any conceivable non-perturbative
contribution.

The renormalization mode of the polarization operator 
$\Pi(Q^2)$  reads  (see, e.g. Ref.~\cite{review})
\beq
\Pi(Q^2/\mu^2,\alpha_s) = Z^{\rm em}  + \Pi_0(Q^2,\alpha_s^0),
{},
\label{renorm:mod}
\eeq
where $\alpha_s = g^2/(4\pi)$ is 
the  strong coupling constant.
Within the  
$\msbar$ scheme \cite{MSbar} 
(here and  below  we are using a convenient combination $a_s = \alpha_s/\pi$)
\beq
Z^{\rm em} = \sum_{1\le j\le i} \left(Z^{\rm em}\right)_{ij} \frac{a_s^{i-1}}{\ep^j} 
{}\, ,
\label{Zem}
\eeq
with the coefficients $\left(Z_{\rm em}\right)_{ij}$ being pure numbers 
and $D=4-2\ep$ standing for the space-time
dimension. 
\ice{
Eq. (\ref{renorm:mod}) can be  naturally deduced from the  connection 
between  $\Pi(Q^2)$
and  the  photon propagator $D_{\mu\nu}(q)$ 
\[
D_{\mu\nu}(q) = g_{\mu\nu}\frac{i}{q^2}\frac{1}{1+e^2\Pi(q^2=-Q^2)}
{}.
\]
}
As a result we arrive at  the following renormalization group (RG) equation
for  the polarization operator  
\beq
\label{rgea}
\left(
\mu^2\frac{\partial}{\partial\mu^2}
 + 
\beta(a_s) 
a_s 
\frac{\partial}{\partial a_s} 
\right)
           \Pi =   \g_{\rm em}(a_s)
{}
\eeq
or, equivalently, ($L_Q = \ln\frac{\mu^2}{Q^2}$) 
\beq
\frac{\prd }{\prd L_Q} \Pi =
\g_{\rm em}(a_s)
-\left(
 \beta( a_s) a_s\frac{\prd }{\prd a_s}
\right) \Pi
\label{rgPi2}
{}.
\eeq
Here
the photon anomalous dimension 
and the  $\beta(a_s)$-function  are  defined in the usual way
\begin{eqnarray}
\g_{\rm em} = \mu^2  \frac{d}{d \mu^2}(Z_{\rm em})-\epsilon
Z_{\rm em} =  
-\sum_{i\geq 0}(i+1)
(Z_{\rm em})_{i1}
a_s ^{i}
{} , 
\label{phpton:anom}
\\
\dmu a_s  = \as \beta(a_s) \equiv
-\sum_{i\geq0}\beta_i a_s^{i+2} 
\label{beta:def}
{}.
\end{eqnarray}
The relation (\ref{rgPi2}) explicitly demonstrates the main computational
advantage of finding first the polarization function $\Pi(Q^2)$
against a direct calculation of $R(s)$ in the case of massless
QCD\footnote{To our knowledge essentially identical observation was
first made in Ref.~\cite{CheKatTka79}}.
Indeed, in order $a_s^n$ the derivative $\frac{\prd }{\prd L_Q} \Pi$
and, consequently, $R(s)$ depends on the very function $\Pi$ which is
multiplied by at least one factor of $a_s$. This means that one needs
to know $\Pi$ up to order $a_s^{n-1}$ only to
unambiguously reconstruct all $Q$-dependent terms in
$\Pi$ to order $a_s^n$, provided, of course, the beta
function and anomalous dimension $\gamma_{\rm em} $  is  known to order 
$a_s^n$. On the   other hand,  the calculation of an anomalous dimension
or a beta-function is known  to be much easier than computing  a correlator
of the same order in the coupling constant (see below).

On specifying the order of $n=3$ we conclude that for computing the
${\cal O} (a_s^3)$ contribution to $R(s)$ one needs to know the
following ingredients:
\begin{itemize}
\item the $\beta$ function to $a_s^2$: it is available through
      independent  calculations of Refs.~\cite{beta2};
\item the polarization function to $a_s^2$ including the constant part
      not depending on $\log Q^2$: 
      the logarithmic contributions were originally obtained in
Refs.~\cite{CheKatTka79,DinSap79CelGon80} while the 
constants were first published in
Refs.~\cite{GorKatLar91,SurSam91}, albeit not in an explicit form;
\item the photon anomalous dimension $\gamma_{\rm em}$ to $\as^3$:
      it is known from works \cite{GorKatLar91,SurSam91} where it was computed with 
      the use of Feynman gauge. 
\end{itemize}        
Thus, in order to check the results of
Refs.~\cite{GorKatLar91,SurSam91} by an independent calculation one
should compute $\Pi$ in order $a_s^2$ (three-loop diagrams) and
$\gamma_{\rm em}$ in order $a_s^3$ (four-loop diagrams).
The corresponding  massless 
Feynman integrals  depend  only on  one 
external momentum, $q$ in our notation,  
and  will be conventionally referred  to as {\em p-integrals} in what follows.

\section{Calculation of p-integrals}

It should be stressed that the first,  three-loop calculation is
definitely much less involved than the second,  four-loop one. There are
three reasons for it.  First, the total number of diagrams is by order
of magnitude less.  Second and most important, there
exists an elaborated algorithm --- the method of integration by parts
of Ref.~\cite{me81a,me81b} --- which allows one to analytically evaluate
divergent as well as finite parts of any three-loop p-integral. Third
and also very important fact is that the algorithm has been neatly and
reliably\footnote{ The package has been extensively tested in various
ways and a probability of the existence of a bug in it seems to be
extremely small. In fact, the calculation we are describing here may
also be considered as another highly non-trivial check of the program.
}  implemented in the language of FORM \cite{Ver91} 
as the package named MINCER in
Ref.~\cite{mincer2}. An important feature of the algorithm is its
ability to evaluate bare dimensionally regulated diagrams. This allows
a convenient two-step evaluation scheme: first calculation of bare
diagrams followed  by a renormalization procedure.  The latter
eventually reduces to a straightforward substitution of the bare
coupling constants expressed through the renormalized ones multiplied
by proper renormalization constants.  Needless to say that such a
substitution is now routinely done with the help of  algebraic
manipulation programs like FORM.

The situation with four-loop diagrams is quite different. At present
there is simply no way to directly compute the  divergent part of a bare
four-loop p-integral, not speaking about its constant part (see in
this connection a discussion in Ref.~\cite{aip2}).  The best what can
be done is the use of the method of Infrared Rearrangement (IRR).
The method is based on on an important observation of
Ref.~\cite{Collins75} that in the $\msbar$-scheme any UV counterterm
has to be polynomial in momenta and masses.  The observation was
effectively employed in Ref.~\cite{Vladimirov80} to simplify
considerably the calculation of UV counterterms.  
The trick  essentially amounts to an appropriate
transformation of the IR structure of Feynman integrals by setting zero some
external momenta and masses (in some cases after some differentiation
is performed with respect to the latter).  As a result the calculation
of UV counterterms is much simplified by reducing the problem to
evaluating  p-integrals.  The method of IRR was ultimately
refined and freed from unessential qualifications 
with inventing a so-called  $R^*$-operation --- a generalization
of the BPHZ $R$-operation to subtract UV and IR divergences ---
in Refs.~\cite{me82,me84}.

The following theorem  has been proven in Ref.~\cite{me84} by the
explicit construction of the corresponding algorithm:
\vglue 0.05cm

\noindent
{\bf
Any UV counterterm for any
(h+1)-loop Feynman
integral can be expressed in terms of pole and finite parts
of some appropriately  constructed (h)-loop p-integrals.
}
\vglue 0.06
cm

\noindent
The above theorem  coupled with the  the integration by parts method
solves  {\em at least in principle} the task
of analytical evaluation of $\g_{\rm em}$ to $a_s^3$.
It should be noted that the $R^*$ -operation is absolutely essential
to prove  the  Theorem, though in most (but not in all) practical
cases one could proceed without it.  However, such a practice, in
fact,  forces  diagram-wise  renormalization mode, what, in turns,
brings down a  heavy penalty of manual  treatment of  hundreds of 
diagrams.

Indeed, in  genuine four-loop calculations the reduction to three-loop 
$p$-integrals  is far from being 
trivial and  includes a lot of  manipulations.
Typical steps  here are
\begin{description}
\item[a]
to reduce the initial  Feynman integral to logarithmically divergent ones via
a proper differentiation with respect  masses and
external momenta;
\item[b]
 to identify  UV
and IR divergent subgraphs of the resulting integral;
\item[c]
to remove  in a recursive way 
the corresponding UV and IR divergences;
\item[d] to  compute   resulting p-integrals.
\end{description}
Among these steps only the calculation of p-integrals can be at
present completely performed by a computer. All others, especially {\bf b}
and {\bf c} are  difficult to computerize.  As a result in  works
\cite{qed-4loop,GorKatLar91,SurSam91} an inherently time-consuming and
error-prone way of manually handling every separate diagram (including
its full UV renormalization  via  $R$-operation) has been
employed\footnote{The calculation  of Ref.~\cite{GorKatLar91} did not  use the 
$R^*$-operation at all while that of Ref.~\cite{SurSam91} employed it only for
a few diagrams.}. 

In present work we will use the full power of the $R^*$-operation to
simplify the steps {\bf b } and {\bf c} so  far  that both UV and IR
renormalizations can be done in a global form and, consequently, can
be simply performed by computer.

\section{IRR in a  global form}

We start from the Dayson--Schwinger equation for the correlator
(\ref{Pi}) written in the bare form\footnote{For simplicity we 
set  the ${}'$t Hooft-Veltman unit of mass $\mu$ equal to $1$ below.}
\beq
 \Pi^0_{\alpha\alpha} (q,a^0_s)  = - \int \mbox{\rm d} p 
  \frac{ (4\pi)^2}{(2\pi)^D}
Tr[\g_\alpha G^0(p+q,a^0_s)\G^0_\alpha(p,q,a^0_s) G^0(p,a^0_s)]
{}.
\label{DSeq:bare}
\eeq
Here $G^0$ and $\G^0_\alpha$ are the full quark propagator and the EM
current vertex function respectively;  below the  
integration with respect to the loop momentum $ p $ with the weight function 
$\frac{ (4\pi)^2}{(2\pi)^D}$  will not be  explicitly displayed.

The renormalized version of (\ref{DSeq:bare}) can be written in two
different but eventually 
equivalent forms: in terms of  renormalization constants or through
\mbox{$R$-operation}. The first form reads
\beq
\Pi_{\alpha\alpha} (q,a_s)  =
Z^{\rm em}q^2(1-D)  - \frac{Z^2_V}{Z_2^2}
Tr[\g_\alpha G^0(p+q,a^0_s)\G^0_\alpha(p,q,a^0_s) \G^0(p,a^0_s)]
{}.
\label{DSeq:ren1}
\eeq
Here $Z_2$ is the quark wave function renormalization constant; 
$Z_V$  is the renormalization constant of the vector   current defined as
\[
[\ovl{\psi}\g_\alpha \psi ] = Z_V/Z_2 \,\,\ovl{\psi^0}\g_\alpha \psi^0
\]
where  the current inside squared   brackets is the
renormalized one. The QED Ward identity implies the equality
$Z_V = Z_2 $, whence  the equivalence of (\ref{renorm:mod}) and 
(\ref{DSeq:ren1}) follows.

The second representation is
\beq
 \Pi_{\alpha\alpha} (q,a_s)  =
Z^{\rm em} q^2(1 -D) - R'
Tr[\g_\alpha G^0(p+q,a_s)\G^0_\alpha(p,q,a_s)  G^0(p,a_s)]
{},
\label{DSeq:ren2}
\eeq
where $R'$  stands for the ``incomplete''  $R$-operation which,  when applied to
a Feynman integral,  subtracts
only all its  UV {\bf sub}divergences not touching  the overall one.

{}From the finiteness of the renormalized correlator we have  two ways
of  finding $Z^{\rm em}$, viz. 
\begin{eqnarray}
Z^{\rm em} &=& - K_\ep \left\{  
\frac{1}{2 D(D-1)}\left(\frac{1}{Z_2} 
\Box_q Tr[\g_{\tilde \alpha} G^0(p+q,a^0_s)\G^0_\alpha(p,q,a^0_s) 
G^0(p,a^0_s)]
\right.
\right.
\\
&+&
\left.
\left.
\frac{\delta Z_V}{Z_2} 
\Box_q Tr[\g_{\alpha} G^0(p+q,a^0_s)\G^0_\alpha(p,q,a^0_s) 
G^0(p,a^0_s)]
\right)
\right\}
{}\, 
\label{Zph1}
\end{eqnarray}
and 
\begin{equation}
Z^{\rm em} =  -K_\ep \left\{  
\frac{1}{2 D(D-1)} 
R' \, \Box_q Tr[\g_{\tilde \alpha} G^0(p+q,a_s)\G^0_\alpha(p,q,a_s) 
G^0(p,a_s)]
\right\}
{}, 
\label{Zph2}
\end{equation}
where $K_\ep f(\ep)$ stands for the singular part of the Laurent
expansion of $f(\ep)$ in $\ep$ near $\ep=0$ and 
$\delta Z_V = Z_V - 1 $.
In Eqs.~(\ref{Zph1}) and
(\ref{Zph2}) we have let a Dalambertian with respect to the external momenta
$q$ act on quadratically divergent diagrams to transform them to the
logarithmically divergent ones.  We also have introduced an auxiliary
mass dependence to a quark propagator --- the one entering into the
``left''  vertex $\gamma_\alpha$--- by making the following change
\beq
\g_\alpha  \to \g_{\tilde \alpha} = \g_\alpha \   p^2/({p^2-m_0^2})
{}.
\label{gamma_tilde}
\eeq

Note, please, that the auxiliary mass dependence has caused somewhat
more complicated structure of UV renormalizations in the right hand
side of Eq.~(\ref{Zph1}). In the case of the second equation corresponding
modifications are taken into account automatically   by $R'$-operation.

Now, to IRR. The idea of the method is quite simple: since the
renormalization constant $Z^{\rm em}$ does not depend on the momentum $q$
dimensionfull one could significantly simplify the calculation of
$Z^{\rm em}$ by nullifying the momentum  Eqs.~(\ref{Zph1}) and
(\ref{Zph2}). The only requirement which must be respected is the
absence of any IR singularities in the resulting bubble
integrals. Unfortunately, a mass introduced to a propagator
is not always sufficient to suppress all IR divergences.
For instance, if $q=0$ then there appear completely massless
tadpoles in the second term on the rhs of Eq.~(\ref{Zph1}).  On the
other hand, in the diagram-wise calculations  of
Refs.~\cite{CheKatTka79,qed-4loop,GorKatLar91, SurSam91}, based on
Eq.~(\ref{Zph2}), it proves possible to tune the position of the
massive propagator for any given diagram 
in such a way to suppress all IR singularities.

Our idea, instead, is to use Eq.~(\ref{Zph1}) supplemented by the
corresponding IR subtractions as prescribed by the $R^*$-operation
formalism \cite{me84}. The problem is facilitated by the fact that, as shown in
Ref~.\cite{me91} the IR counterterm constants for a given diagram may
be determined in terms of some properly chosen combination of UV ones.
The only remaining task  is to write the IR subtractions in a global way.
We have done it with the  following result:
\begin{eqnarray}
Z^{\rm em} &=& - K_\ep \left\{  
\frac{1}{2 D(D-1)} 
\Box_q Tr[\g_{\tilde \alpha} G^0(p+q,a^0_s)\G^0_\alpha(p,q,a^0_s) 
G^0(p,a^0_s)]|_{{}_{\displaystyle q=0}}
\right.
\nonumber
\\
&-& 
\left.
\frac{1}{Z_2}
\frac{1}{4 D}Tr[\delta\G^0_{\tilde\alpha}(0,0,a^0_s)\g_\alpha ] Z^{\rm em}
-
\frac{\delta Z_V }{Z_2} Z^{\rm em}
\right\}
{}\, .
\label{final_eq}
\end{eqnarray}

Several comments are in order regarding this formula.

First,  Eq.~(\ref{final_eq}) is,  rigorously speaking,
applicable as it stands  only to the so-called non-singlet diagrams, that
is to those where both EM currents belong to one and the same quark
loop.  The four singlet diagrams, violating this requirement, appear
first  in  order $a_s^3$.  A full derivation of (\ref{final_eq}) with
modifications necessary to include singlet case will be presented elsewhere.
                                
Second, by $\delta\G^0_{\tilde\alpha}(p,q,a^0_s)$
we denote the vertex function of the electromagnetic current
with the tree contribution removed. The ``tilde'' atop the index 
$\alpha$  again  means that  in every diagram the quark propagator 
entering to the vertex $j_\alpha$ is softened at small momenta by
means of the auxiliary mass $m_0$  according to Eq.~(\ref{gamma_tilde}).
The bare coupling constant  $a_s^0$ is to be understood as 
$a_s = Z_a a_s$,  with  $Z_a$ being 
the coupling constant renormalization constant.  

Finally,  an inspection of (\ref{final_eq}) immediately shows that,
in order to find the ${(n+1)}$-loop  correction to $Z^{\rm em}$, 
one  needs only to  know the renormalization constants $Z_2$ and
$Z^{\rm em}$  to order $a_s^n$ as well as    the bare Green functions 
\beq
 G^0(p,a^0_s), \ \ 
\frac{\partial}{\partial q_\beta }
[\G^0_\alpha(p,q,a^0_s)]|_{{}_{\displaystyle q=0}} 
, \ \ 
\Box_q [\G^0_\alpha(p,q,a^0_s)]|_{{}_{\displaystyle q=0}}, 
\ \ 
\delta\G^0_{\tilde\alpha}(0,0,a^0_s)
\label{functions}
{}
\eeq
up to (and including) $n$-loops,  that is to order  $(a_s^0)^n$.
Thus, we have achieved  our aim and   obtained   a general formula for 
$Z^{\rm em}$  in terms of bare \mbox{$p$-integrals} with explicitly resolved 
UV and IR subtractions.

\section{Results and discussion}

We have computed with the program MINCER the unrenormalized three-loop
Green functions (\ref{functions}) as well as the quark wave function
renormalization constant $Z_2$ to order $a_s^3$.  The calculations
have been performed in the general covariant gauge with the gluon
propagator $(g_{\mu\nu} - \xi \frac{q_\mu q_\nu}{q^2})/q^2$.  We have
also taken into account the singlet diagrams as well as extra diagrams
with some of virtual quarks replaced by colour octet neutral fermions.
In the minimal supersymmetric standard model such a fermion known as
gluino appears as the superpartner of the gluon. The total
calculational time with an DEC workstation exceeds 200 hours for the
general gauge; for the Feynman one it is reduced to about 20 hours.

Then we have used Eqs.~(\ref{final_eq}) and (\ref{rgPi2})  
to find $\g_{em}$  to order $a_s^3$. We have used the following 
values for the coefficients of the beta-function  in QCD with gluinos
\cite{betaQ8}
\begin{equation} 
\label{beta:coeff}
\begin{array}{lll} \displaystyle
\beta_0
& = & \displaystyle
\frac{1}{4}\left[
\frac{11}{3} C_A - \frac{4}{3} T n_f 
- \frac{2}{3} C_A n_{\tilde g} \right]
{},
\\ \displaystyle
\beta_1
& = & \displaystyle
\frac{1}{16}\left[
\frac{34}{3} C_A^2 - 4 C_F T n_f
- \frac{20}{3} C_A T n_f 
- \frac{16}{3} C_A^2 n_{\tilde g}
\right]
{}.
\end{array}
\end{equation}
\ice{
In[6]:= Collect[(-betaQ8)/.{no->ng/2},a]

           11 ca   2 ca ng   4 nf tr
Out[6]= a (----- - ------- - -------) + 
             3        3         3
 
              2        2
      2  34 ca    16 ca  ng   20 ca nf tr
>    a  (------ - --------- - ----------- - 4 cf nf tr)
           3          3            3
}
\noindent
Here  $C_A$ and $C_F$  
are the Casimir operators of the adjoint and
quark (defining) representations of the colour group; $T$ is the
normalization of the trace of generators of quark representation
$Tr(t^a t^b) = T \delta^{ab}$; $ n_f $ is the number of quark
flavours; $d[R]$ is the dimension of the quark representation of the
colour group and $n_{\tilde g}$ is the number of neutral colour octets which we
take either zero or one.

Our results  for $\g_{em}$ and $R(s)$ read 
\begin{eqnarray}
\lefteqn{\dsp \gamma^{em} =  
 d[R]{}
 \sum_f Q_f^2 \left\{ \rule{.0mm}{0.5cm}
 \frac{4}{3}
\, {+}\, a_s  \,C_F 
\right.}
\nonumber
\\
&{}&
\rule{0.1cm}{0cm}
\left.
{+} a_s^2 \left[C_F^2
\left(
-\frac{1}{8}\right)
{+} \,C_F \,C_A 
\left(
 \frac{133}{144}\right)
{+} \,C_F \,T \,n_f 
\left(
-\frac{11}{36}\right)
{+} \,C_F \,C_A  \,n_{\tilde g}
\left(
-\frac{11}{72}\right)
\right]
\right.
\nonumber
\\
\nonumber
\\{}
\nonumber 
\\{}
\nonumber 
\break
+a_s^3\left[\rule{.0mm}{0.5cm}
\right. 
&{}&
C_F^3
\left(
-\frac{23}{32}\right)
{+} C_F^2\,C_A 
\left(
\frac{215}{216} 
-\frac{11}{18}  \,\zeta(3)
\right)
{+} \,C_F C_A^2
\left(
\frac{5815}{15552} 
+\frac{11}{18}  \,\zeta(3)
\right)
\nonumber\\
&{+}& C_F^2\,T \,n_f 
\left(
-\frac{169}{216} 
+\frac{11}{9}  \,\zeta(3)
\right)
{+} \,C_F \,C_A \,T \,n_f 
\left(
-\frac{769}{3888} 
-\frac{11}{9}  \,\zeta(3)
\right)
\nonumber\\
&{+}& \,C_F  T^2 \, n_f^2
\left(
-\frac{77}{972}\right)
{+} \,C_F \,C_A  \,n_f \,T \,n_{\tilde g}
\left(
-\frac{77}{972}\right)
{+} C_F^2\,C_A  \,n_{\tilde g}
\left(
\frac{19}{216} 
+\frac{1}{9}  \,\zeta(3)
\right)
\nonumber\\
&{+}& \,C_F C_A^2 \,n_{\tilde g}
\left(
-\frac{4495}{7776} 
-\frac{1}{9}  \,\zeta(3)
\right)
{+} \,C_F C_A^2 n_{\tilde g}^2
\left(
-\frac{77}{3888}\right)
\left.
\left.
\rule{.0mm}{0.5cm}
\right]
\right\}
\nonumber
\\
+ a_s^3
&{}&
\!\!\!\!\!\!\!\!\!\!\!\! 
\left(\sum_f Q_f \right)^2
\frac{d_{abc}d_{abc}}{256}
\left(
\frac{176}{9} - \frac{128}{3}\zeta(3)
\right)
{},
\label{gamma:em}
\end{eqnarray}
\begin{equation}
\begin{array}{l}
\dsp
R(s) = d[R]\sum_f Q_f^2\left\{1
+ 
a_s(\mu)
r_1
+
a^2_s(\mu)
\left[
s_2 
+
\lgm\frac{\mu^2}{s} \left(
s_1\beta_0 
   \right)
\right]
\right.
\\ 
\dsp 
\rule{3.4cm}{0cm}
\left.
+
a^3_s(\mu)
\left[s_3
+
\lgm\frac{\mu^2}{s} \left( 2 s_2\beta_0
+s_1 \beta_1
   \right)
+
\lgm^2\frac{\mu^2}{s} \left(
 s_1\beta_0^2 
   \right)
\right]
\right\}
\\ 
\dsp
\rule{3.4cm}{0cm}
+ a_s^3
\left(\sum_f Q_f \right)^2
\frac{d_{abc}d_{abc}}{1024}
\left[
\frac{176}{3} -  128\zeta(3)
\right]
{},
\end{array}
\label{RS0}
\end{equation}
\begin{eqnarray}
r_1 &{=}&  
  \,C_F 
\left[
 \frac{3}{4}\right]
{},
 \ \ \
 r_2 = 
C_F^2
\left[
-\frac{3}{32}\right]
{+} \,C_F \,C_A 
\left[
\frac{123}{32} 
-\frac{11}{4}  \,\zeta(3)
\right]
{+} \,C_F \,T \,n_f 
\left[
-\frac{11}{8} 
+  \,\zeta(3)
\right]
\nonumber\\
&{}&
\rule{4.65cm}{0cm}
{+} \,C_F \,C_A  \,n_{\tilde g}
\left[
-\frac{11}{16} 
+\frac{1}{2}  \,\zeta(3)
\right],
\nonumber\\
{r_3} &=& 
{} C_F^3
\left[
-\frac{69}{128}\right]
{+}\,  C_F^2\,C_A 
\left[
-\frac{127}{64} 
-\frac{143}{16}  \,\zeta(3)
+\frac{55}{4}  \,\zeta(5)
\right]
\nonumber\\
{+} &{}& \!\!\!\!\!\!\!\!\!\! 
\,C_F C_A^2
\left[
\frac{90445}{3456} 
-\frac{121}{576}  \pi^2
-\frac{2737}{144}  \,\zeta(3)
-\frac{55}{24}  \,\zeta(5)
\right]
{+}\,  C_F^2\,T \,n_f 
\left[
-\frac{29}{64} 
+\frac{19}{4}  \,\zeta(3)
-5  \,\zeta(5)
\right]
\nonumber\\
{+} &{}& \!\!\!\!\!\!\!\!\!\! 
\,C_F \,C_A \,T \,n_f 
\left[
-\frac{485}{27} 
+\frac{11}{72}  \pi^2
+\frac{112}{9}  \,\zeta(3)
+\frac{5}{6}  \,\zeta(5)
\right]
{+}\,  \,C_F  T^2 \, n_f^2
\left[
\frac{151}{54} 
-\frac{1}{36}  \pi^2
-\frac{19}{9}  \,\zeta(3)
\right]
\nonumber\\
{+} &{}& \!\!\!\!\!\!\!\!\!\! 
C_F \,C_A  \,n_f \,T \,n_{\tilde g}
\left[
\frac{151}{54} 
-\frac{1}{36}  \pi^2
-\frac{19}{9}  \,\zeta(3)
\right]
+
C_F^2\,C_A  \,n_{\tilde g}
\left[
\frac{9}{16} 
+\frac{13}{8}  \,\zeta(3)
-\frac{5}{2}  \,\zeta(5)
\right]
\nonumber\\
{+} &{}& \!\!\!\!\!\!\!\!\!\!  \,C_F C_A^2 \,n_{\tilde g}
\left[
-\frac{33767}{3456} 
+\frac{11}{144}  \pi^2
+\frac{251}{36}  \,\zeta(3)
+\frac{5}{12}  \,\zeta(5)
\right]
{+} \, C_F C_A^2 n_{\tilde g}^2
\left[
\frac{151}{216} 
-\frac{1}{144}  \pi^2
-\frac{19}{36}  \,\zeta(3)
\right],
\nonumber\\
\label{r3 }
\end{eqnarray}
where  
$ d^{abc} =2  Tr(\{ t^a t^b \} t^c) $.

We observe that neither $\g_{em}$ no $R(s)$ depend on the gauge fixing
parameter $\xi$ as it must be. If $n_{\tilde g}$ is set to zero
then $R(s)$ is in complete agreement with the results of 
Refs.~\cite{GorKatLar91,SurSam91}.  
For the  standard QCD  $SU_c(3)$
colour group values  $C_F= 4/3, \, C_A = 3, \, T = 1/2$
and $d^{abc}d^{abc} = 40/3$ we get for $R(s)$ with $\mu^2 = s$
\begin{eqnarray}
\lefteqn{R(s) = 
3\sum_f Q_f^2 \left\{ \rule{.0mm}{0.5cm}
\right.
1
{+} a_s 
{+} a_s^2 
\left[
\frac{365}{24} 
-11  \,\zeta(3)
-\frac{11}{12}  \,n_f 
+\frac{2}{3}  \,\zeta(3) \,n_f 
-\frac{11}{4}  \,n_{\tilde g}
+2  \,\zeta(3) \,n_{\tilde g}
\right]}
\nonumber\\
&{+}& a_s^3 
\left[ \ \ 
\frac{87029}{288} 
-\frac{121}{48}  \pi^2
-\frac{1103}{4}  \,\zeta(3)
+\frac{275}{6}  \,\zeta(5)
-\frac{7847}{216}  \,n_f 
+\frac{11}{36}  \pi^2 \,n_f
\Break
\phantom{+ a_s^3 }
+\frac{262}{9}  \,\zeta(3) \,n_f 
-\frac{25}{9}  \,\zeta(5) \,n_f 
+\frac{151}{162}  \, n_f^2
-\frac{1}{108}  \pi^2 \, n_f^2
-\frac{19}{27}  \,\zeta(3) \, n_f^2
\Break
\phantom{+ a_s^3 }
-\frac{32903}{288}  \,n_{\tilde g}
+\frac{11}{12}  \pi^2 \,n_{\tilde g}
+\frac{277}{3}  \,\zeta(3) \,n_{\tilde g}
-\frac{25}{3}  \,\zeta(5) \,n_{\tilde g}
+\frac{151}{27}  \,n_f  \,n_{\tilde g}
\Break
\phantom{+ a_s^3 }
-\frac{1}{18}  \pi^2 \,n_f  \,n_{\tilde g}
-\frac{38}{9}  \,\zeta(3) \,n_f  \,n_{\tilde g}
+\frac{151}{18}  n_{\tilde g}^2
-\frac{1}{12}  \pi^2 n_{\tilde g}^2
-\frac{19}{3}  \,\zeta(3) n_{\tilde g}^2
\right]
\left.
\rule{.0mm}{0.5cm}
\right\}
\nonumber
\\ 
&{}&
{}
\rule{.1cm}{0cm}
+ 
\rule{0.5cm}{0cm}
a_s^3
\left(\sum_f Q_f \right)^2
\left(
\frac{55}{72} - \frac{5}{3}\zeta(3)
\right)
\label{Rqcd }
\end{eqnarray}
or, in the numerical form, 
\bea
R(s) &=& 3\sum_f Q_f^2 \left\{
1 + a_s +
 a_s^2 \left( 1.98571 - 0.115295 n_f - 0.345886 n_{\tilde g}
\right)
\right.
\nonumber
\\
&+& a_s^3
\left(
-6.63694 - 1.20013 n_f - 0.00518 n_f^2  - 2.85053 n_{\tilde g} 
\right.
\nonumber
\\
&{}&
\left.
\left.
\rule{0.8cm}{0cm}
-0.03107 n_f n_{\tilde g} - 0.04661 n_{\tilde g}^2 
\right)
\right\}
\nonumber
\\
&-&
\as^3 \left(\sum_f Q_f\right)^2
1.2395
{}.
\label{Rnum1}
\eea
\ice
{                    2
Out[14]= 1. + as + as  (1.98571 - 0.115295 n_f - 0.345886 n_{\tilde g}) + 
 
       3                                       
>    as  (-6.63694 - 1.20013 n_f - 0.00517836 n_f^2  - 2.85053 n_{\tilde g} -

>       0.0310701 n_f n_{\tilde g} - 0.0466052 n_{\tilde g}^2 )
}

At last, for the phenomenologically  relevant case of $n_f = 5 $  we obtain
\beq
R(s) = \frac{11}{3}
\left[
1 + a_s  a_s^2(1.409 - 0.346 n_{\tilde g}) + 
+a_s^3 (-12.805 - 3.006 n_{\tilde g}- 0.0466 n^2_{\tilde g})
\right]
{}.
\label{Rnum2}
\eeq

\noindent 
Thus, the ${\cal O}(\alpha_s^3)$ gluino contribition to $R(s)$ has the
same (negative) sign as in the $\alpha_s^2$ order.  We should also
add  that for a meaningful phenomenological discussion of the
gluino contribution to $R(s)$ one should also take into account the
running of the coupling constant in the next-next-to leading order.
This requires the knowledge of the gluino contribution to the
three-loop coefficient $\beta_2$ which, to our knowledge, is not
yet available in the literature (the purely QCD contribution to 
$\beta_2$ is  known from \cite{TarVlaZha80,Larin:betaQCD}.

To summarize: we have suggested a new convenient way to compute the UV
renormalization constant of the correlator of vector quark currents. Our
final  formula (\ref{final_eq}) directly expresses the
constant in terms of unrenormalized p-integrals, with all UV and IR
subtractions being implemented in a global form.  The formula is
 useful in carrying out completely automatic 
calculations. In our previous work \cite{gssq} similar formula
has been obtained for the case of the correlator of scalar currents.

Using the formula and the FORM version of MINCER \cite{mincer2} we
have computed the ${\cal O}(\alpha_s^3)$ correction to $R(s)$ in pQCD
including light gluino. In a particular case of the standard QCD we
have reproduced the result of Refs.~\cite{GorKatLar91,SurSam91}.  This
gives also an extra support for the non-accidental nature of the
findings of Ref.~\cite{BroadKataev93}.  An important  feature of
our calculation was the use of the general covariant gauge, which has
allowed us to demonstrate for the first time the gauge independence of
$R(s)$ at  ${\cal O}(\alpha_s^3)$.
\vskip0.3cm

 \noindent
{\Large{\bf Acknowledgments}}
\vskip0.3cm

I would like to thank G.~Farrar, G. Gagabadze, M.~Je\.{z}abek,
A.~Kataev, J.~K\"uhn, B.~Kniehl, S.~Larin, M.~Steinhauser,
A.~Pivovarov and  V.~Rubakov for stimulating and useful discussions. I am
specially grateful to J. Vermasseren and S. Larin for providing me
with the FORM package MINCER as well useful advice about its features.
I would like to thank S. Jadach for the occasion to present the
results of this paper at the Cracow International Symposium on
Radiative Corrections, 1-5 August 1996.

I am deeply thankful to the Institute of Theoretical Particle Physics
of the Karlsruhe University and the theoretical group of the
Max-Plank-Institute of  Physics and Astrophysics for the warm
hospitality. 

This work was supported by  
INTAS under Contract INTAS-93-0744.

\sloppy
\raggedright
\def\app#1#2#3{{\it Act. Phys. Pol. }{\bf B #1} (#2) #3}
\def\apa#1#2#3{{\it Act. Phys. Austr.}{\bf #1} (#2) #3}
\def\lhc{Proc. LHC Workshop, CERN 90-10}
\def\npb#1#2#3{{\it Nucl. Phys. }{\bf B #1} (#2) #3}
\def\plb#1#2#3{{\it Phys. Lett. }{\bf B #1} (#2) #3}
\def\prd#1#2#3{{\it Phys. Rev. }{\bf D #1} (#2) #3}
\def\pR#1#2#3{{\it Phys. Rev. }{\bf #1} (#2) #3}
\def\prl#1#2#3{{\it Phys. Rev. Lett. }{\bf #1} (#2) #3}
\def\prc#1#2#3{{\it Phys. Reports }{\bf #1} (#2) #3}
\def\cpc#1#2#3{{\it Comp. Phys. Commun. }{\bf #1} (#2) #3}
\def\nim#1#2#3{{\it Nucl. Inst. Meth. }{\bf #1} (#2) #3}
\def\pr#1#2#3{{\it Phys. Reports }{\bf #1} (#2) #3}
\def\sovnp#1#2#3{{\it Sov. J. Nucl. Phys. }{\bf #1} (#2) #3}
\def\jl#1#2#3{{\it JETP Lett. }{\bf #1} (#2) #3}
\def\jet#1#2#3{{\it JETP Lett. }{\bf #1} (#2) #3}
\def\zpc#1#2#3{{\it Z. Phys. }{\bf C #1} (#2) #3}
\def\ptp#1#2#3{{\it Prog.~Theor.~Phys.~}{\bf #1} (#2) #3}
\def\nca#1#2#3{{\it Nouvo~Cim.~}{\bf #1A} (#2) #3}

\end{document}